2016

# Systematic Mapping Protocol

HAVE SYSTEMATIC REUSE BENEFITS BEEN TRANSFERRED TO REAL-WORLD SETTINGS?

FINAL VERSION: 2016/08/02

José Luis Barros Justo. Universidad de Vigo. ESEI. España
Fernando Óscar Pinciroli. Faculty of Informatics and Design, Champagnat University, (5501) Godoy Cruz, Argentina
Santiago Matalonga. Universidad ORT. Uruguay
Marco Aurelio Paz González (student). Universidad de Vigo. ESEI. España
Nelson Martínez Araujo (student). Universidad de Vigo. ESEI. España

# Contents

# Introduction

Since its first appearance in the NATO Conference in 1968 software reuse has been defined as a way to develop new software (or modify existing one) based on previously developed assets. By using software reuse the developers should avoid the need to "reinvent the wheel" every time they start a new project, therefore reducing development time. As some "pieces" were reused several times the chance that bugs remain undetected reduces, so the quality of reused assets (and the software built upon them) increase with every new use. These two main claims (reduced development time and increased quality) were paraphrased as: reduce development and maintenance effort, reduce bugs, decrease faults rate and so. The ultimate benefit, from the industry point of view, can be stated as: reduce costs.

There are a lot of published works both in Journals and Conferences which keep making these claims but, little evidence (real data) in the context of industrial applications has been offered. We are interested in answer the big question: Are these claims about the benefits of software reuse in industrial context real or just a myth?

This document details the planning phase of a Systematic Mapping Study. Our goal is to identify and to understand the benefits that the software engineering community has reported on the application of the different reuse strategies in industrial context, by building a general picture (map) containing: the claimed benefits, the data supporting those claims, the industry's domains and the reuse strategy employed.

# Planning

## Research goal and Questions

The objective of this work is to map current research that describes the benefits of the existing reuse techniques in Software Engineering and their impacts in real world settings.

Therefore, we describe the research goal (RG) as:

- Identify and classify the benefits that software reuse has delivered in real-world settings

The main research question (MRQ) for the mapping study is formulated as:

- What evidence has been reported that software reuse achieves its purposes of delivering benefits in real-world settings?

This research question is broad enough to allow us to slice the research space in two dimensions: topic research space and publication space. The topic research space is mutable, it depends strongly on the research topic and, therefore the research questions (RQs) are subject to change in every systematic study. On the other hand, the questions in the publication space (Publication Questions, PQs from now on) can be fixed beforehand, as they are very similar in many systematic reviews [Kitchenham B.; Chartes et al. 2007], (Brereton P., 2011) and (Petersen,et al., 2015).

## Topic research space

RQ1: Which benefits have been reported? (frequency and evolution in time)

RQ2: Were the benefits assessed? If yes, how?

RQ3: Were the benefits monetarizing? (yes/no)

RQ4: Type of real-world setting? (Industry's domain)

RQ5: Type of reuse? (Systematic/Planned reuse versus Ad Hoc/Opportunistic reuse)

RQ6: Reuse process? (Reuse-for (Domain Engineering, SPL), Reuse-with (CBD, MBD, COTS))

RQ7: Duration? (Time frame of the reported experience)

RQ8: Were threats to validity reported? (yes/no)

## Publication space

PQ1: Top venues: Conferences/Journals; and their evolution (published works/year)

PQ2: Publications per year (its evolution, no matter the venue)

PQ3: Top-cited papers (H-index)

PQ4: Active researchers (counting cites)

PQ5: Researcher's affiliation (Academic/Industry)

PQ6: Active countries (main author affiliation)

## Search strategy

We use four independent search strategies: 1) automatic search on four different online databases, 2) forward and backward snowballing with an initial set of papers from our Related Work section, 3) manual search on selected conferences that usually publish reuse related research (International Conference on Software reuse, 1996-2015; ICCBSS (International Conference on Commercial-off-the-Shelf (COTS)-Based Software Systems, 2002-2007; International Conference on Composition-Based Software Systems, 2008; and SPLC (Software Product Line Conference, 2004-2014), and 4) manual search on selected authors (rank by productivity) by visiting their personal web pages or retrieving their publications profile.

The automatic search was carried out using the following databases:

- ACM Digital Library
- IEEE Xplore
- Scopus
- Web of Science

### Search String

We applied the PICOC Structure suggested by (Kitchenham and Charters, 2007) to identify keywords that could be used to build up the search string (see Annex I. Evolution of the Search String).

Where:

- **Population**: Software Reuse
  *"software reuse", reusability, reusable, "domain engineering", "software product line", SPL, "component-based", CBD, "model-based", MBD, MDD, commercial-of-the-shelf, COTS, framework, pattern*
- **Outcomes**: Benefits of reuse
  *quality, productivity, reliability, portability, time, cost, time-to-market, performance*
- **Context**: Industry or Academia
- **Limits**: White literature, between 1968 and 2015. When available, we limit the results to those in the subject area ("Computer Science/Software Engineering").

Search string was tailored to the databases as detailed in the following table:

Table 1   Search strings

| Database | Search string |
|---|---|
| ACM DL | keywords.author.keyword:("software reuse" reusab* "domain engineering" "software product line" SPL "component-based" CBD "model-based" MBD MDD commercial-of-the-shelf COTS framework pattern) AND<br>keywords.author.keyword:(quality productivity reliability portability time cost time-to-market performance) AND<br>keywords.author.keyword:(industry organization firm business compan*) |
| IEEE Xplore | (("Index Terms":"software reuse" OR "Index Terms":reusab* OR "Index Terms":"domain engineering" OR "Index Terms":"software product line" OR "Index Terms":"component-based" OR "Index Terms":COTS) AND (Search_Index_Terms:quality OR "Index Terms":productivity OR "Index Terms":reliability OR "Index Terms":portability OR "Index Terms":"time-to-market" OR "Index Terms":performance) AND ( "Index Terms":industry OR "Index Terms":organization OR "Index Terms":firm OR "Index Terms":business OR "Index Terms":compan*) AND ( "Index Terms":"software engineering")) |
| SCOPUS | KEY[1]("software reuse" OR reusab* OR "domain engineering" OR "software product line" OR SPL OR "component-based" OR CBD OR "model-based" OR MBD OR MDD OR commercial-of-the-shelf OR COTS OR framework OR pattern) AND<br>KEY(quality OR productivity OR reliability OR portability OR time OR cost OR time-to-market OR performance) AND<br>KEY(industry OR organization OR firm OR business OR compan*) AND<br>(DOCTYPE(ar) OR DOCTYPE(cp) ) AND<br>(PUBYEAR > 1967 AND PUBYEAR < 2016) AND<br>KEY("software engineering") AND<br>( LIMIT-TO(SUBJAREA,"COMP" ) ) |
| Web of Science | #1: TS[2]=("software reuse" OR reusab* OR "domain engineering" OR "software product line" OR SPL OR "component-based" OR CBD OR "model-based" OR MBD OR MDD OR commercial-of-the-shelf OR COTS OR framework OR pattern)<br>#2: TS=(quality OR productivity OR reliability OR portability OR time OR cost OR time-to-market OR performance)<br>#3: TS=(industry OR organization OR firm OR business OR compan*)<br>#4: #1 AND #2 AND #3 AND TS=("software engineering")<br> #4 Refined by: RESEARCH AREAS: ( COMPUTER SCIENCE ) AND DOCUMENT TYPES: ( ARTICLE )<br>Timespan: 1968-2015. |

## Forward and backwards snowballing

Snowballing was conducted according to the guidelines by (Wohlin, 2014). Citations count as well as the search for the referenced works was performed using SCOPUS.

## Inclusion/Exclusion Criteria

The following inclusion/exclusion criteria were applied to all four described search strategies.

---

[1]KEY is a combined field that searches the AUTHKEY, INDEXTERMS, TRENDNAME and CHEMNAME fields
[2] TS field includes: title + abstract + key terms fields

Inclusion Criteria:

1. Source must be in English.
2. Source must be peer-reviewed.
3. Source must be published between 1968 and December 2015.
4. Source must be relevant to the software engineering domain.
5. Source must discuss one of:
    a. Software reuse (in any of its possible techniques)
    b. The experience in a real world setting (Industry)
    c. The benefit from applying the reuse technique.

Exclusion criteria

1. Negation of all the Inclusion criteria.
2. Source is excluded if it describes the reuse technique in an academic setting

## Data Extraction form

In addition to bibliographic information of each source, the following fields were extracted from each source:

- Benefits: Description of the resulting benefits as per the wording of the authors.
- Assessed: An indicator (yes/no) to signal if the authors have evaluated the previous benefit(s).
- Method: A description of the method used to evaluate the benefit(s).
- Monetized: An indicator to signal if the benefits were monetized.
- Domain: Industrial domain as described on the source. As per the wording of the authors, a vlue of "Not Reported" as added if this information could not be extracted from the source.
- Type of Software reuse: Either Systematic or Opportunistic.
- Reuse process applied: One of "Domain Engineering"; "Software Product Lines"; "Component Based Development"; "Model Based Development"; "Commercial Of The Shelf". These typesof reuse process are defined in IEEE 1517:2010
- Threats to validity: a description of the threats to validity discussed in the source.
- Type of study: One of "Quasi-experiment"; "Case Study"; "Survey"; "Solution Proposal"; "Experience Report"; "Expert Opinion". The definition for each category is described in Annex II.

The following image presents an example of our data extraction sheet template.

|   | A | B | C | D | E | F | G | H | I | J | K | L |
|---|---|---|---|---|---|---|---|---|---|---|---|---|
| 1 |   |   |   |   |   |   | Research Space |   |   |   |   |   |
| 2 | Paper_ID | Author_year | RQ1 | RQ2 | RQ2 - How? | RQ3 | RQ4 | RQ5 | RQ6 | RQ7 | RQ8 | Study type |
| 3 | P1 |   |   |   |   |   |   |   |   |   |   |   |
| 4 | P2 |   |   |   |   |   |   |   |   |   |   |   |
| 5 | P3 |   | quality | Yes |   | Yes | Aeronautics | Systematic | DE |   | Yes | Quasi-experiment |
| 6 | P4 |   | productivity | No |   | No | Banking | Opportunistic | SPL |   | No | Case Study |
| 7 | P5 |   | reliability |   |   |   | Insurance |   | CBD |   |   | Survey |
| 8 | P6 |   | portability |   |   |   | Health |   | MBD |   |   | Solution Proposal |
| 9 | P7 |   | development- |   |   |   | Computers (Hardware) |   | COTS |   |   | Experience Report |
| 10 | P8 |   | development- |   |   |   | Computers (Software) |   |   |   |   | Expert Opinion |
| 11 | P9 |   | maintenance- |   |   |   | Automotive |   |   |   |   |   |
| 12 | P10 |   | time to marke |   |   |   |   |   |   |   |   |   |
| 13 | P11 |   | performance |   |   |   |   |   |   |   |   |   |
| 14 | P… |   |   |   |   |   |   |   |   |   |   |   |

Figure 1  DEF for Research Space questions

For questions in the Publication Space the template was configured as follows:

|   | A | N | O | P | Q | R | S |
|---|---|---|---|---|---|---|---|
| 1 |   |   |   | Publication Space |   |   |   |
| 2 | Paper_ID | PQ1 (Venue) | PQ2 (Year) | PQ3 (# Cites) | PQ4 (First au | PQ5 (A/I) | PQ6 (Country) |
| 3 | P1 | Name of the |   | SCOPUS |   | (A)cademic |   |
| 4 | P2 | Conference or |   |   |   | (I)ndustry |   |
| 5 | P3 | the Journal |   |   |   |   |   |
| 6 | P4 |   |   |   |   |   |   |
| 7 | P5 |   |   |   |   |   |   |

Figure 2  DEF for Publication Space questions

# Conclusions

We have strictly followed the guidelines published by [Petersen_2015] to develop a SMS. As the whole team adhered to these guidelines to build up the protocol presented in this document we think the execution phase of the proposed protocol will be repeatable, and that internal threats to validity can be mitigated.

## Acknowledgments

We further wish to acknowledge the work done by: Ania Cravero at Chile (Universidad de La Frontera) and Marcela Genero at Spain (Universidad de Castilla-La Mancha) in reviewing this protocol.

# Annex I. Evolution of the Search String

We detail the evolution of the search string for the Related Work section. Once the final search string was built we suppressed the Intervention facet to adapt the string for the search of primary works for the systematic mapping study.

Situation at 25/02/16:   **For the Related Work section**

We used PICOC, instead of PICO, and some synonyms for the main terms.

Population: is software reuse, as a subarea of software engineering. We select the following terms as synonyms: "software reuse", reusability, reusable, "domain engineering", "software product line", "component-based development", "model-based development" and "commercial-of-the-shelf". The last five terms correspond to software reuse processes.

Intervention: was set to mapping studies or literature reviews. Since we are interested in retrieving as many studies as possible we do not include the "systematic" term in Intervention, making the original search broader. The following terms were selected: mapping, review, evidence.

Comparison: not applicable

Outcomes: different benefits mentioned in software reuse publications. The selected terms were: quality, productivity, reliability, portability, "development-time", "development-cost", "maintenance-cost", "time-to-market" and performance. They all come from the standard IEEE Std. 1517/2010.

Context: real-world settings. Terms included were: industry, industrial, firm, company, business and organization (and their plural forms).

01/04/2016:

Population: "software reuse", reusability, "domain engineering", "software product line", "component-based development", "model-based development", "commercial-of-the-shelf"

Intervention: review, study, mapping, survey, evidence

Outcomes: productivity, quality, reliability, portability, "development time", "development cost", "maintenance cost", "time-to-market"

Context: industry, industrial, industries, organization, firm, business, company, companies

Key terms from other sources:

- Mogagheghi_2007: software reuse, review, quality, productivity, evidence; reliability, reusable software
- V. Bauer_2013: industrial, internal, maintenance, software quality, software reuse
- Krueger_1992: software reuse
- Mili_1995: software reuse, reusable components
- Frake_2005: software reuse, finance, software productivity, software quality, software reusability
- Lim_1994: industrial economics, personnel, quality assurance, software reuse, software productivity, software quality
- Morisio_2002: empirical study, software reuse, survey; software reusability
- Selby_2005: software reuse, empirical study; software reusability
- Haefliger_2008: knowledge reuse, software reuse; software reusability, software reliability, code reuse

The search string should be composed as:

Population AND Outcomes AND Context AND Intervention

But if we apply Context terms the retrieved set of papers is too limited and, well-known papers such as the one from Mohagheghi, were not found. So at the end we eliminate Context (all the terms) from the search string.

Some examples of searches:

ACM Digital Library (Full-text Collection):

"query": { keywords.author.keyword:("software reuse" reusab* "domain engineering" "software product line" "component-based development" "model-based development" "commercial-of-the-shelf") AND keywords.author.keyword:(productivity quality reliability portability "development time" "development cost" "maintenance cost" "time-to-market") AND keywords.author.keyword:(review mapping evidence) }

7 results (37 results if Searched The ACM Guide to Computing Literature, but we'll use only the Full-text Collection as we are interested only in papers which are full-text accessible)

IEEE Xplore (Metadata Only):

| Index Terms | Combined field which allows users to search the Author Keywords, DOE Terms, IEEE Terms, INSPEC Terms, Mesh Terms, and PACS Terms. |
|---|---|

("Index Terms":"software reuse" OR reusab* OR "domain engineering" OR "software product line" OR "component-based development" OR "model-based development" OR "commercial-of-the-shelf") AND ("Index Terms":productivity OR quality OR reliability OR portability OR "development time" OR "development cost") AND ("Index Terms":review OR mapping OR evidence)

("Index Terms":"software reuse" OR "Index Terms":reusab* OR "Index Terms":"domain engineering" OR "Index Terms":"software product line" OR "Index Terms":"component-based development" OR "Index Terms":"model-based development" OR "Index Terms":"commercial-of-the-shelf") AND ("Index Terms":productivity OR "Index Terms":quality OR "Index Terms":reliability OR "Index Terms":portability OR "Index Terms":"development time" OR "Index Terms":"development cost") AND ("Index Terms":review OR "Index Terms":mapping OR "Index Terms":evidence)

53 results    (47 Conference papers, 5 journal articles, 1 Standard)    (IEEE impose a maximum of 15 search terms)

(08/04/2016): **Final version of search string**

**ACM DL:**

"query": { keywords.author.keyword:("software reuse" reusab* "domain engineering" "software product line" "component-based" "model-based" COTS) AND keywords.author.keyword:(productivity quality reliability portability "development time" "development cost" "maintenance cost" "time-to-market" performance) AND keywords.author.keyword:(review mapping  evidence) }

8 results (too few, so we relax the condition changing keywords.author.keyword: by recordAbstract)

"query": { keywords.author.keyword:("software reuse" reusab* "domain engineering" "software product line" "component-based" "model-based" COTS) AND recordAbstract:(productivity quality reliability portability "development time" "development cost" "maintenance cost" "time-to-market" performance) AND recordAbstract:(review mapping  evidence) }

97 results

**IEEE Xplore:**

("Index Terms":"software reuse" OR "Index Terms":reusab* OR "Index Terms":"domain engineering" OR "Index Terms":"software product line" OR "Index Terms":"component-based" OR "Index Terms":"model-based" OR "Index Terms":"commercial-of-the-shelf") AND ("Index Terms":productivity OR "Index Terms":quality OR "Index Terms":reliability OR "Index Terms":portability OR "Index Terms":"development time" OR "Index Terms":"development cost" OR "Index Terms":"maintenance cost" OR "Index Terms":"time-to-market" OR "Index Terms":performance) AND ("Index Terms":review OR "Index Terms":mapping OR "Index Terms":evidence)

No results (search terms exceed 15) so we shortened the search string:

("Index Terms":.QT.software reuse.QT. OR "Index Terms":reusab* OR "Index Terms":.QT.domain engineering.QT. OR "Index Terms":.QT.software product line.QT. OR "Index Terms":.QT.component-based.QT. OR "Index Terms":.QT.model-based.QT. OR "Index Terms":COTS) AND ("Index Terms":productivity OR "Index Terms":quality OR "Index Terms":reliability OR "Index Terms":portability OR "Index Terms":.QT.development cost.QT. OR "Index Terms":"maintenance cost" OR "Index Terms":performance) AND ("Index Terms":review OR "Index Terms":mapping OR "Index Terms":evidence)

164 results

**SCOPUS:**

(KEY ("software reuse" OR reusab* OR "domain engineering" OR "software product line" OR "component-based" OR "model-based" OR COTS)) AND ( KEY ( productivity OR quality OR reliability OR portability OR "development time" OR "development cost" OR "maintenance cost" OR "time-to-market" OR performance) ) AND ( KEY ( review OR mapping OR evidence) ) AND ( LIMIT-TO(SUBJAREA,"COMP" ) )

126 results (3 from 2016; 73 conference papers; 45 articles; 6 review; 2 in Press)

**WoS:**

#1: TS=("software reuse" OR reusab* OR "domain engineering" OR "software product line" OR "component-based" OR "model-based" OR COTS)

#2: TS= (productivity OR quality OR reliability OR portability OR "development time" OR "development cost" OR "maintenance cost" OR "time-to-market" OR performance)

#3: TS=(review OR mapping OR evidence)

#4: TS="software engineering"

#5: #1 AND #2 AND #3 AND #4      63 results

**Duplicates**

| Duplicates | ACM DL | IEEE Xplore | SCOPUS | WoS |
|---|---|---|---|---|
| ACM DL | 3 | | 5 | 1 |
| IEEE Xplore | | 2 | 11 | |
| SCOPUS | | | | 7 |
| Total | 9 | 13 | | 7 |

29 papers to remove. We did not remove papers from SCOPUS.

The final string to apply to the search of primary works was (in SCOPUS format):

KEY ("software reuse" OR reusab* OR "domain engineering" OR "software product line" OR SPL OR "component-based" OR CBD OR "model-based" OR MBD OR MDD OR commercial-of-the-shelf OR COTS OR framework OR pattern) **AND**

KEY (quality OR productivity OR reliability OR portability OR time OR cost OR time-to-market OR performance) **AND**

KEY (industry OR organization OR firm OR business OR compan*) **AND**

(DOCTYPE(ar) OR DOCTYPE(cp) ) **AND**

(PUBYEAR > 1967 AND PUBYEAR < 2016) **AND**

KEY ("software engineering") AND (LIMIT-TO(SUBJAREA,"COMP" ) )

# Annex II. Study type classification taxonomy

To identify, catalogue, and analyse empirical work assessing reuse, we follow [1, 2, 3, 4]. We classified empirical studies of reuse into the following categories:

***Quasi-Experiment***. In a quasi-experiment, one or more characteristics of a controlled experiment are missing, such as strict experimental control and/or randomization of treatments and subject selection. This is typical in industrial settings. The researcher has to enumerate alternative explanations for observed effects one by one, decide which are plausible, and then use logic, design, and measurement to assess whether that might explain any observed effect [1].

***Case Study***. A case study is an empirical inquiry that investigates a contemporary phenomenon within its real-life context, especially when the boundaries between phenomenon and context are not clearly evident. In a case study, all of the following exist: research questions, propositions (hypotheses), units of analysis, logic linking the data to the propositions, and criteria for interpreting the findings [2]. Observational studies are either case studies or field studies. Case studies focus on a single project, while multiple projects are monitored in a field study, maybe with less depth. Case studies may also involve analysis of historical data [3].

***Survey***. A survey consists of structured or unstructured questions given to participants. The primary means of gathering qualitative or quantitative data in surveys are interviews or questionnaires [1]. Structured interviews (qualitative surveys) with an interview guide, investigate rather open, and qualitative research questions with some generalization potential. Quantitative surveys with a questionnaire contain mostly closed questions [4].

**Solution Proposal**. This type of paper proposes a solution technique and argues for its relevance, without a full-blown validation. The technique must be novel, or at least a significant improvement of an existing technique. A proof-of-concept may be offered by means of a small example, a sound argument, or by some other means.

***Experience Report***. An experience report is similar to a case study, but it does not have the same level of controls or measures. It is retrospective, generally lacks propositions, may not answer how or why phenomena occurred, and often includes lessons learned [4]. In this chapter, we combine example applications with experience reports because most papers had features of both. An example application consists of "authors describing an application and providing an example to assist in the description, but the example is 'used to validate' or 'evaluate' as far as the authors suggest [20]," but without the rigor of a formal case study.

***Expert Opinion***. An expert opinion provides some qualitative, textual, opinion-oriented evaluation. It is "based on theory, laboratory research, or consensus [3]." These expert opinions assess processes, strategies, approaches, theoretical models, policies, curriculum, or technology that may or may not allude to full-scale evaluation or empirical studies. Often such articles are based on experience, observations, and ideas proposed by the author(s).